\documentclass{llncs}

\newcommand{\citeneed}[1]{}
\newcommand{\keywords}[1]{\par\addvspace\baselineskip
\noindent\keywordname\enspace\ignorespaces#1}

\def\land{\wedge}
\def\convex{{\bf X}}

\def\ellip{\mathcal{E}}
\def\reals{{\bf R}}
\def\sympos{{\bf SP}}

\usepackage{makeidx}  
\usepackage{graphicx}
\usepackage{subfigure}
\usepackage{url}

\begin{document}
\mainmatter

\title{Using Ellipsoidal Domains to Analyze Control Systems
Software\thanks{This material is based upon work supported by the National
Science Foundation under Grant 0615025 and by NASA under grant NNX08AE37A}}
\author{Fernando Alegre\and Eric Feron\and Santosh Pande}
\institute{Georgia Institute of Technology, Atlanta, Georgia, USA}
\maketitle

\begin{abstract}
We propose a methodology for the automatic verification of safety properties
of controllers based on dynamical systems, such as those typically used
in avionics. In particular, our focus is on proving stability properties
of software implementing linear and some non-linear controllers. We
develop an abstract interpretation framework that follows closely the
Lyapunov methods used in proofs at the model level and describe the
corresponding abstract domains, which for linear systems consist of
ellipsoidal constraints. These ellipsoidal domains provide abstractions
for the values of state variables and must be combined with other domains
that model the remaining variables in a program. Thus, the problem of
automatically assigning the right type of abstract domain to each variable
arises. We provide an algorithm that solves this classification problem
in many practical cases and suggest how it could be generalized to more
complicated cases.
We then find a fixpoint by solving a matrix equation,
which in the linear case is just the discrete Lyapunov equation.
Contrary to most cases in software analysis,
this fixpoint cannot be reached by
the usual iterative method of propagating constraints until saturation
and so numerical methods become essential. Finally, we illustrate our
methodology with several examples.
\keywords{analysis, control systems, software verification, abstract
          interpretation, Lyapunov methods}
\end{abstract}

\section{Introduction}

In this paper we propose a new methodology for the automatic verification of controllers
based on dynamical systems, such as those typically used in avionics.
One of the most important properties that a controller needs to have is stability, i.e.,
the ability to stay close to the reference value at all times\footnote{
This definition applies to open-loop systems, which are the focus of our paper.
Closed-loop stability needs a joint model of controller and plant.}.
Stability is a critical aspect
of the design of a controller, and a guarantee of stability is often a requirement for certification
of safety-critical controllers. While testing a design is usually helpful to perfect it, it is not
enough to ensure the correct behavior. Thus, a model of the design is usually created, either
by hand or by some
software tool, and the model is checked for the desired properties.

At its core, a linear controller for a dynamical system is typically modeled as a recursive equation
\begin{equation}\label{eq:typical}
\begin{array}{c}
x_0 = 0 \\
x_{n+1} = f(x_n,y_n) \\
u_{n+1} = g(x_{n+1})
\end{array}
\end{equation}
where the variable $y_n$ is the $n$-th input read from the physical system (called the \emph{plant})
and $u_n$ is the $n$-th output sent to the plant. Data interchange occurs periodically, and there
may be synchronization problems if the exchange does not happen at regular time intervals, but in
this paper we assume that timing issues are not critical and do not affect the behavior of either
the controller or the plant. Outputs are computed as a function of the inputs and of some internal
state represented by the variable $x_n$. An equation similar to equation \ref{eq:typical} is also
used to model the plant.

Current engineering practice is to use software tools to assist in the design of the models. These
tools, such as Mathworks' Simulink and Real Time Workshop \cite{mathworks:rtw}, also provide
\emph{autocoders} that automatically generate implementations of controllers directly from the models.
The C source code generated by autocoders may be either generic or specialized for a certain processor,
depending on the options chosen by the user, and sometimes very simple models may be translated into
thousands of lines of cryptic C code.

As software implementations become increasingly more widespread, the need to guarantee
their proper behavior becomes more important. While this problem can be seen as merely
checking that a piece of code correctly implements the specifications, there is a non-trivial
amount of work in matching such specifications, usually in the form of an abstract automaton,
with the corresponding variables and functions in the
code. Such a \emph{template matching} approach has been carried out in some cases
\cite{feretA4static}, but it is not clear how to generalize it to more complex cases and whether
this approach scales well. Since autocoders are just special-purpose compilers, we could
reverse-engineer them to
come up with the matching. Alternatively, we could also either treat the problem as a decompilation
problem or enhance the autocoder with the generation of a proof, thus converting an autocoder into
a credible compiler \cite{rinard99credible}. All of these approaches depend heavily on the
particular autocoder, thus limiting their applicability without serious amount of effort whenever
different autocoders, or even different versions of an autocoder are considered. Thus, it is
desirable to use a methodology for verification that depends as little as possible on either the
particularities of an autocoder or the specific patterns in the code.

Much progress has been done recently in automatic analysis of models of controllers
\cite{henzinger97hytech}\cite{alurA3progress}, especially when the state space is finite
(and thus susceptible to model checking approaches) or can be reduced to a finite case.
These model checking techniques are most suitable for hardware implementations of controllers.
However, software implementations need additional considerations, as the software itself
introduces many aspects not captured by the original model, such as pointer indirection,
enlarged state spaces due to auxiliary variables, and, most importantly, greater complexity
of the underlying model, which is the main reason for the shift from hardware to software.
While software controllers still have a finite number of states, for all practical purposes
this number must be considered infinite or intractable.

Thus, our only choice left to validate the implementation is whether to analyze the specification
and rely on the correctness of the autocoder\footnote{Unlike the case for C compilers
\cite{leroyA6formal}, to our knowledge there is no ongoing research on certification
of any autocoder.}
or to analyze the implementation independently of the specification.
We chose the latter approach because
it is obvious that the lower the level, the closer to the actual executable, and thus the
more useful the verification would be. Moreover, high-level specification languages such as Matlab
do not have
a clearly specified semantics, and the proprietary single implementation provided is the only
authoritative reference. In contrast, lower level languages, such as C, while still ambiguous
and implementation-dependent in some details, are largely well-defined, and their semantics is
well known. Thus, instead of attempting to prove correctness of the implementation with respect
to some specification, we will show that it is possible
to prove that the generated code satisfies the desired property,
which in our case is stability\footnote{As mentioned earlier, in this paper we focus on open-loop
controller stability.}.

Our methodology consists of
building a framework based on software analysis, and in particular abstract interpretation,
and using this analysis to synthesize a proof that can be independently checked with a simpler proof
checker. Our technique resembles the Proof-Carrying Code technique \cite{necula97pcc} and has
similar advantages. While code analysis and abstract interpretation techniques have previously
been applied to similar problems \cite{blanchetA3static}, our technique is easier to generalize
to more complex cases, due in part to the explicit separation of code analysis and proof synthesis,
and in part to our use of a particular type of abstract domain.
It is also one of the first attempts to approach
the verification of stability of software controllers in a systematic way, so that even though
the examples provided in this paper a somewhat limited, a path towards more complex cases is
clearly indicated.

The main contributions of this paper are the following:
\begin{enumerate}
\item In order to perform static analysis of a controller implementation, we introduce
a characterization of variables, which we call roles, that can distinguish between variables
comprising a dynamical state and variables containing dynamical parameters,
and give an algorithm to classify all variables according to their respective roles.
\item We introduce an abstract domain for variables in state roles that describes a
controller loop as a transformation of a multi-dimensional ellipsoid.
\item We show how local propagation rules along with the use of a constraint solver give
way to a global proof of stability.
\end{enumerate}

The rest of the paper is organized as follows. Section \ref{sec:background} summarizes the
theoretical foundations of our work. Section \ref{sec:analyzing} describes our methodology.
Section \ref{sec:synthesizing} explains how our method is applied.
Section \ref{sec:related} positions our
method with respect to similar approaches. Finally, section \ref{sec:conclusion} summarizes
our work
and explains how to extend our it in several ways.

\section{Background}\label{sec:background}

Automatic verification often relies on using constraints from within a family for which
an apriori guarantee of convergence can be provided. Constraints can be expressed
either as a logical formula or as an equivalent set relation. For example, this
code fragment
\begin{verbatim}
i=0; while(i<10) { A[i] = 0; i++; }
\end{verbatim}
can be annotated as follows:
\[
\begin{array}{l}
\{ \} \verb|i=0;| \{i=0 \land I(i) \} \\
\verb|while(i<10) {| \{ G(i) \land I(i) \} \verb|A[i] = 0;i++;| \{ G(i-1) \land I(i) \} \verb|}| \\
\{i=10 \land I(i) \} \\
\mbox{where } G(i) = 0\le i < 10
\mbox{ and } I(i) = \forall i',0\le i'< i \Rightarrow A[i']=0
\end{array}
\]

We made explicit the dependency of formulas $G$ and $I$ on their free variables, but it is still
obvious that they are just the usual guard and invariant of the loop. However, coming up with an
automatic decision procedure to infer them is not obvious at all. One recent approach is given in
\cite{gulwaniA8lifting}, where the authors build an abstract domain based on guarded universally
quantified formulas and provide an algorithm for joins and meets based on under-approximating
the guard operations. Alternatively, we can use a parametric abstract domain approach to derive
the invariant. The key observation is that the statement \verb|A[i] = 0;| can be interpreted as
a projection. Without loss of generality or concern for floating point issues, which are
orthogonal to our goal here, let us assume that \verb|A| is a real valued array of dimension 10.
Then the statement \verb|A[i] = 0;| projects the value $A\in \reals^{10}$ onto the subspace
orthogonal to the $i$-th coordinate axis: $\mathcal{F}_i A = A - (A\cdot e_i)e_i$.
If, before executing
the statement $A$ was in some set $D(i)$, then after executing it, we will have
\[
\{ A \in D(i) \} \verb|A[i] = 0;|
\{ A \in \mathcal{F}_i D(i) \}
\]
where $\mathcal{F}_i D(i) = \{x\in\reals^{10} : \exists y\in D(i), x = \mathcal{F}_i y\}$.

Let us further assume that there is a function $g$ such that $\mathcal{F}_i D(i) \subset D(g(i))$.
It is easy to see that these assumptions can only hold if $g(i) = i+1$ and the recursion
$D(0)=\reals^{10}, \mathcal{F}_i D(i) \subset D(i+1)$ has a solution. An obvious solution in this case
is $D(i) = (\prod_{j=0}^{i-1} \mathcal{F}_j) \reals^{10}$.
Thus, the sole assumption that the variable \verb|A| lives
in a parametric abstract domain is enough to generate an equation for the domain.
Finding a solution means that the corresponding abstract interpreter exists,
and thus any analysis that uses it is valid.

This methodology for verification divides the process into three separate steps:
first use of analysis to propose suitable abstract domains,
then use of some domain knowledge to solve the domain equations,
and finally use of abstract interpretation to validate the required properties.

It is tempting to interpret the analysis phase as some kind compilation where the statements of a
program are translated into abstract statements implementing an abstract interpreter. This abstract
interpreter is usually constructed by Kleene iteration on partial interpreters with the hope that
a fixpoint is reached in a finite number of steps \cite{cousot77abstract}. Here we take an optimistic
approach and jump directly to the fixpoint, provided that it can be expressed as a function
of a finite set of parameters and that it is possible to determine the values of those parameters
by some other means.

Many common abstract domains are actually parametric abstract domains according to this definition:
\begin{definition}
A {\bf parametric abstract domain} $(D_f,X,L)$ is given by a predicate $f(x,p)$ with free variables
$x$ and $p$, a set $X$ and a lattice $L$ so that for all $p\in L$, the subsets
\[D_f(p) = \{x \in X : f(x,p) \}\]
satisfy
\begin{enumerate}
\item $D_f(p) \cup D_f(q) \subset D_f(p \sqcup q)$
\item $D_f(p) \cap D_f(q) \supset D_f(p \sqcap q)$
\end{enumerate}
\end{definition}

Parametric abstract domains are useful when the program to be verified is compatible:
\begin{definition}
A program $S$ with $X$-valued variables $x$ is {\bf compatible} with a parametric abstract domain
$(D_f,X,L)$
if every statement $\sigma$ in $S$
has a transfer function $g_\sigma$ such that
\[\{x \in D_f(p)\} \sigma \{x \in D_f(g_{\sigma}(p))\}\] holds.
\end{definition}

Thus, compatible parametric abstract domains allow us to translate a program $S$ into a
composition of functions $g_S$ so that $S$ admits fixpoints of the form $D_f(p)$
where the parameters $p$ are just solutions to $p=g_S(p)$.

\subsection{Related abstract domains}

The simplest abstract domain consists of providing upper and lower bounds to each scalar 
variable independently. This is known as the rectangular abstract domain. While simple to
manipulate, this domain does not capture any interdependence between variables and is prone
to grow until the superset of reachable values becomes the entire state space. Nevertheless,
the rectangular abstract domain has been used with partial success for verification of control
models \cite{preussig99reachability}.

Relational domains assume a particular type of relationship between the variables. The
polyhedral domain arises when relationships are assumed to be linear or affine. In this
case, overapproximations of the reachable set in the form of polyhedrons are sought.
While this method is much more precise,
the computational burden is very high, which makes it unsuitable for
problems with more than a handful of variables \cite{henzingerA1some}.

A trade-off between complexity and precision is provided by the octagonal domain, in which
affine relationships between variables are limited to those with coefficients $+1$ or $-1$
only \cite{mineA6octagon}. This domain has seen recent success in verification of not only models
but also implementations, and it remains a useful domain in many circumstances. However,
for the particular problem we are interested in, the ellipsoidal domain is still a superior choice.

\subsection{The Ellipsoidal Domain}

The ellipsoidal abstract domain for the analysis of dynamical systems is
ubiquitous and has been met with much theoretical and practical success.
The central reason behind this fact is that, for all practical purposes,
existence of an invariant ellipsoid is equivalent to dynamical system
stability: To be more precise, Lyapunov's stability theory
\cite{vidyasagar92nonlinear}
states that the linear system defined by the recursion
\[
x_{k+1} = Ax_k
\]
is stable (all trajectories converge to zero) if and only if there exists an
ellipsoidal invariant. The possibility to approximate any smooth nonlinear
system
\[
x_{k+1} = f(x_k)
\]
in the vicinity of a given equilibrium by a linear system
provides a straightforward justification
to apply the search for invariant ellipsoids to such systems as well.

We now present two formulations of an ellipsoid which we will use
interchangeably below.

\begin{definition}
In the direct formulation, an ellipsoid in $\reals^n$ is the set
\begin{displaymath}
\ellip(P) = \{x \in \reals^n : x^t P x \le 1\}
\end{displaymath}
where the matrix $P \in \reals^{n \times n}$
is symmetric and positive semi-definite.
\end{definition}

\begin{definition}\label{def:reverse}
In the reverse formulation, an ellipsoid in $\reals^n$ is the set
\begin{displaymath}
\ellip^\dag(P) =
\{x \in \reals^n : \left(
\begin{array}{cc} 1 & x^t \\ x & P \end{array}
\right)\ge 0 \}
\end{displaymath}
where the matrix $P \in \reals^{n \times n}$ is symmetric and
positive semi-definite.
\end{definition}

The direct formulation includes degenerate ellipsoids which are cylindrical in
some dimensions, i.e., the set is unbounded along directions corresponding
to eigenvectors of $P$ with null eigenvalue, and the reverse formulation includes
degenerate ellipsoids which are flat in some
dimensions, i.e., the corresponding semi-axes have null length.

A direct application of Schur decomposition shows that if the matrix $P$ is invertible, then
the direct formulation
and the reverse formulation are equivalent, and $\ellip^\dag(P) = \ellip(P^{-1})$. Informally,
a degenerate case in the latter would correspond to having some infinite eigenvalues in the former.

\section{Analyzing autocoded models}\label{sec:analyzing}

\subsection{Identification of state variables}

In order to apply our propagation rules, we need to classify variables into 3 groups:
state variables, indices and parameters. In this section we describe our assumptions and
explain the algorithm to perform this partitioning. However, we make no guarantees that
any program can be partitioned in such a way.
Only programs whose variables can be partitioned can be analyzed by our method.

A typical dynamical system uses variables in two different ways. Some variables represent the
unknown internal states of the system while the rest are the known parameters. The system is
usually specified as differential (or difference if discrete) equations on the unknown states
with some initial condition. An approximate solution to these equations is usually needed, since
the output of the controller depends on them. Parameters may or may not be constant and may or
may not depend on the internal state.

The main assumption in this paper is that
parameters do not depend on the internal state.
In that case, we can characterize state as those variables that satisfy these conditions:

\begin{enumerate}
\item State changes with time from an initial given value.
\item Future values of the state are calculated from its current value, the current inputs
and the parameters\footnote{this is the formal definition of state normally used in dynamical systems
theory}
\item Values of parameters are independent of any future, past or present value of the state
\item Output values are calculated from the state and the parameters.
\end{enumerate}

Additionally, a C program introduces a third type of variable due to the non-atomic nature of
arrays, i.e., since arrays are accessed component-wise, we also need variables to hold the
relevant indices.

In order to translate these conditions into a static analysis of a piece of code, we need
to use a concept similar to liveness, which we will call persistence.
The persistent range of a variable is similar to the live range, but it
persists across redefinitions. Formally, a variable is persistent between two
nodes $n_1$ and $n_2$ in the control flow graph if $n_2$ is an immediate successor
of $n_1$ and the variable is live-out at $n_1$ and live-in at $n_2$. In this analysis, subscripted
variables, consisting of a base name and one or more index expressions, are stripped off the
subscripts and treated as if just the base was present, so for example
the variable \verb|x| will persist across the assignment \verb|x[i] = a[i]*x[j-1]|.
The persistence ranges of a variable are then defined as the connected components
of the control flow graph induced by the persistence relationship.

If a variable has more than one persistence range, we rename it to a different unique name
in each of the ranges, so that there is only one persistence range for each variable.
This ensures that a single role can be assigned to each variable. We then use the following
algorithm to classify the variables:

\begin{enumerate}
\item Any variable that appears in a subscript expression is marked as index.
\item Index variables are then ignored in the rest of this analysis.
\item A dependency graph is formed according to these rules:
      \begin{enumerate}
      \item A variable in the left-hand side of an assignment depends on all the variables
            in the right-hand side.
      \item A variable dominated by an assertion node depends on the variables in the assertion.
      \item Dependency is a transitive relationship.
      \end{enumerate}
\item Any input variable is marked as state.
\item Any variable that depends on itself is marked as state.
\item Finally, all variables not marked as state or index are marked as parameters.
\end{enumerate}

This algorithm can be made finer grained so that
independent components of the state are treated separately, but for this paper the coarse
version presented suffices.

\subsection{Propagation of constraints}

The most important rule describes how assignments propagate ellipsoids forward. We
specialize the generic forward assignment rule to some important cases occurring in control
systems.
In particular,
when the expression is a linear transformation of the state $x = Ax$ and the pre-condition is
an ellipsoidal constraint $x \in \ellip^\dag(P)$, we can give a simpler formula for the
post-condition.

\begin{lemma}\label{lemma:linear}
An assignment of a vector to a linear transformation of itself given by a matrix $A$ is
associated to the Hoare triple
\begin{equation}\label{eq:linear}
\{ x \in \ellip^\dag(P) \} \verb| x = A*x; | \{ x \in \mathcal{R}(A) \cap \ellip^\dag(Q) \}
\end{equation}
where $\mathcal{R}(A) = \{x: \exists y, x = Ay \},Q=APA^t$.
\end{lemma}

Lemma \ref{lemma:linear} follows from the implication $P \ge 0 \Rightarrow \forall M, MPM^t \ge 0$
when the choice $M = \left(\begin{array}{cc}1 & 0 \\ 0 & A\end{array}\right)$ is made.
Even though the postcondition in rule (\ref{eq:linear}) is not the
strongest possible, it is strong enough for our purposes.

Rule (\ref{eq:linear})
is actually a rule schema because it represents
many different situations found in the code.
Important special cases of this rule concern the introduction of new variables, dropping
of dead variables and initialization of variables by a constant, which can be
represented by the choices
$A=\left(\begin{array}{c}I \\ 0\end{array}\right)$,
$A=\left(\begin{array}{cc}I & 0\end{array}\right)$ and
$A=\left(\begin{array}{cc}I & 0\\ 0 & 0\end{array}\right)$
respectively, leading to Hoare triples such as

\begin{equation}\label{eq:addzero}
\begin{array}{c}
\{ x\oplus y \in \ellip^\dag(R_1)\}
\verb| y = 0; |
\{ x\oplus y \in \ellip^\dag(R_2)\} \\
R_1 = \left(\begin{array}{cc}P & 0 \\ 0 & Q\end{array}\right)
R_2 = \left(\begin{array}{cc}P & 0 \\ 0 & 0\end{array}\right)
\end{array}
\end{equation}

For a more elaborate example,
consider the assignment

\begin{verbatim}
x[i] = x[i] + a[i][j]*x[j];
\end{verbatim}

where variables $i$ and $j$ are integer indices, array $x$ is the state and array $a$ is constant
and not part of the state. This assignment can be expressed as

\begin{displaymath}
x \gets (I + a_{ij} E_{ij}) x
\end{displaymath}
where $I$ is the identity matrix and $E_{ij}$ is a matrix with all zeros except for the $ij$-th
cell, which is one. In that case, the following triple holds:

\begin{displaymath}
\begin{array}{c}
\{\left(\begin{array}{c}x\\y\end{array}\right) \in
\ellip^\dag\left(\begin{array}{cc}P & R^t \\ R & Q\end{array}\right) \} \\
\verb| x[i] = x[i] + a[i][j]*x[j]; | \\
\{\left(\begin{array}{c}x\\y\end{array}\right) \in
\ellip^\dag\left(\begin{array}{cc}TPT^t & TR^t \\ RT^t & Q\end{array}\right) \} \\
\end{array}
\end{displaymath}
where
$T = I + a_{ij} E_{ij}$.

While the propagation rules for ellipsoids in direct form are more complicated, it is
sometimes not possible to avoid using them.

Rules in the direct formulation will often need to be expressed in terms of 
of convex combinations of constraints. In order to simplify our notation, we introduce
the following definition.

\begin{definition}
The $n$-dimensional convex-combinator space $\convex^n$ is the set
\begin{displaymath}
\convex^n = \{ \vec{\lambda} \in [0,1]^n : \sum_{i=1}^n \lambda_i = 1 \}
\end{displaymath}
\end{definition}

It is also worth noting that rules in the direct formulation are often
universally quantified in terms of some parameter.

\subsubsection{Intersection of ellipsoids}

When a variable satisfies a finite number of ellipsoidal constraints $\{x\in\ellip(P_i)\}$, we
want to
give a single ellipsoidal constraint that overapproximates the intersection
$\{x\in\cap_i \ellip(P_i)\}$. In the direct formulation, this is easily stated.

\begin{lemma}\label{lemma:intersection}
The intersection of $n$ ellipsoids can be overapproximated by any convex combination
of them as follows:
\begin{displaymath}
\frac
{\begin{array}{ccc}\{x \in \ellip(P_1)\} & \dots & \{x \in \ellip(P_n)\}\end{array}}
{\forall \lambda \in \convex^n, \{ x \in \ellip(\sum_{i=1}^n \lambda_i P_i) \}}
\end{displaymath}
\end{lemma}

\subsubsection{Cartesian product}\label{sub:cartesian}

Lemma \ref{lemma:intersection} has an important special case when distinct constraints involve
disjoint vectors. Then, the sum itself becomes just the quadratic form associated to
the Cartesian product of each individual ellipsoid.
Therefore, we have the following rule
schema, in which we assume that variables $x_1, \dots, x_n$ are disjoint:

\begin{equation}\label{eq:product}
\frac
{\begin{array}{ccc}\{x_1 \in \ellip(P_1)\} & \dots & \{x_n \in \ellip(P_n)\}\end{array}}
{\forall \lambda \in \convex^n, \{ x_1 \oplus \dots \oplus x_n \in \ellip(P_\lambda)\}}
\end{equation}
\begin{equation}
P_\lambda =
\left(\begin{array}{cccc}
\lambda_1 P_1 & 0 & \dots & 0 \\
0  & \lambda_2 P_2 & \dots & 0 \\
\dots & \dots & \dots & \dots \\
0 & 0 & \dots & \lambda_n P_n \end{array}
\right)
\end{equation}

This rule needs to be applied after some external input is read. The input is assumed to satisfy
some ellipsoidal constraint which may need to be combined with the corresponding constraint
satisfied by the internal state. The rule is also needed if the input or the initial constraint
satisfy a rectangular bound, as then each co-ordinate variable satisfies a 1-dimensional
ellipsoidal constraint.

\subsubsection{Projection}

The conceptual opposite of introducing a new variable with its associated constraint is forgetting
some variable and transforming the joint constraint into a constraint involving only the remaining
variables. While this is just a particular case of lemma \ref{lemma:linear} and thus always
doable in the reverse formulation, it is not always possible to do in the direct formulation, as we
need some additional regularity conditions.

\begin{lemma}\label{lemma:projection}
Let $x \in \reals^n$ and $y \in \reals^m$ be two disjoint vectors that satisfy the joint constraint
\begin{displaymath}
\{x\oplus y \in \ellip\left(\begin{array}{cc}P & R^t \\ R & Q\end{array}\right)\}
\end{displaymath}
\begin{enumerate}
\item if $Q$ invertible, then the vector $x$ satisfies $\{x \in \ellip(P - R^t Q^{-1} R)\}$.
\item if $R$ is null, then the vector $x$ satisfies $\{x \in \ellip(P)\}$.
\end{enumerate}

\end{lemma}


The semantic rule associated with lemma \ref{lemma:projection} can be applied
whenever a variable goes dead or we no longer need to track it. This usually happens after its
value is written to some output channel, or when the value of an auxiliary variable is reverted
back to a main variable.

\subsubsection{Assignments in direct formulation}

As was the case for projection, rules for assignments in the direct formulation are more
complicated than their reverse form counterparts and not always applicable, as they
depend on some regularity conditions.

\begin{lemma}
An assignment of a vector to a linear transformation of itself given by an invertible matrix $A$
is associated to the Hoare triple
\begin{equation}\label{eq:lineardirect}
\{ x \in \ellip(P) \} \verb| x = A*x; | \{ x \in \ellip(A^{-t}PA^{-1}) \}
\end{equation}
\end{lemma}

Note that if $P$ is also invertible, then rule (\ref{eq:lineardirect}) is
equivalent to rule (\ref{eq:linear}).

We also have an alternative rule for copy operations of the form $y = x$,
and in general for assignments of
the form $y = Ax$ where $A$ is an $n \times m$ matrix and $x$ and $y$
are disjoint. In this case, we can add
to a precondition of the form $x^t Px \le 1$ a multiple of the vanishing term
$(y - Ax)^2$ to arrive to the following lemma.

\begin{lemma}
Let $A$ be an $n\times m$ matrix. Then, the following triple holds:
\begin{displaymath}
\begin{array}{c}
\{x \in\ellip(P) \} \verb| y = A*x; | 
\{\forall \lambda\in\reals, x\oplus y \in \ellip(Q_\lambda)\} \\
Q_\lambda =
\left(\begin{array}{cc}P & 0 \\0 & 0\end{array}\right)
+ \lambda\left(\begin{array}{c}A^t \\ -I\end{array}\right)
\left(\begin{array}{cc}A & -I\end{array}\right)
\end{array}
\end{displaymath}

If $P$ is invertible, then the following triple also holds:

\begin{displaymath}
\begin{array}{c}
\{x \in\ellip^\dag(\hat{P}) \} \verb| y = A*x; | 
\{\forall \epsilon\in\reals, x\oplus y \in \ellip^\dag(\hat{Q})\} \\
\hat{Q} =
\left(\begin{array}{cc}
\hat{P} & \hat{P}A^t \\ A\hat{P} & A\hat{P}A^t + \epsilon I
\end{array}\right)
\end{array}
\end{displaymath}
where $\hat{P} = P^{-1}$.
\end{lemma}

\subsubsection{Nonlinear expressions}

Let us introduce now a simple non-linear
expression\footnote{this expression is commonplace in control systems theory}
suitable for our analysis as given in the
following lemma.

\begin{lemma}\label{lemma:globlips}
Let $f:\reals^n \to \reals$ be a function satisfying the global
condition $\forall x\in \reals^n, |f(x)| \le |x|$. Then, an assignment
of the form \verb|u = f(x);| can be associated to the triple
\begin{displaymath}
\begin{array}{c}
\{ x \in \ellip(P) \}
\verb| u = f(x); | \\
\{\exists \mu_0 > 0,\forall \mu \in [0,\mu_0],x \oplus u \in \ellip(Q)\} \\
Q =
\left(\begin{array}{cc}P-\mu I & 0 \\ 0 & \mu I\end{array}\right)
\end{array}
\end{displaymath}
In addition, if $P$ is invertible, then the following triple also holds:
\begin{displaymath}
\begin{array}{c}
\{ x \in \ellip^\dag(\hat{P}) \}
\verb| u = f(x); | \\
\{\exists \epsilon_0 > 0,\forall \epsilon \ge \epsilon_0,x \oplus u \in \ellip^\dag(\hat{Q})\} \\
\hat{Q} =
\left(\begin{array}{cc}
-\epsilon\hat{P}(\hat{P}-\epsilon I)^{-1} & 0 \\ 0 & \epsilon I
\end{array}\right)
\end{array}
\end{displaymath}
where
$\hat{P} = P^{-1}$.
\end{lemma}

Lemma \ref{lemma:globlips} follows from adding the term $\mu(u^2-|x|^2)$,
which is non-positive whenever $\mu$ is non-negative, to the ellipsoidal
constraint in the pre-condition.


\begin{figure}
$\{ x \in \ellip^\dag(P) \}$ \\
\verb|y = c*x;| \\
$\{ x\oplus y \in \ellip^\dag
\left(\begin{array}{cc}P & Pc^t \\ cP & cPc^t + \epsilon_y I\end{array}\right)
= \ellip^\dag (P_1(\epsilon_y))
\}$ \\
\verb|u = f(y);| \\
$\{ x\oplus y\oplus u\in\ellip^\dag\left(\begin{array}{cc}
-\epsilon_u P_1(\epsilon_y)[P_1(\epsilon_y)-\epsilon_u I]^{-1} & 0 \\
0 & \epsilon_u I
\end{array}\right)
= \ellip^\dag (P_2(\epsilon_y,\epsilon_u))
\}$ \\
\verb|x = x + b*u;| \\
$\{
x\oplus y\oplus u\in\ellip^\dag(AP_2(\epsilon_y,\epsilon_u)A^t),
A = \left(\begin{array}{ccc}I & 0 & b \\ 0 & I & 0 \\ 0 & 0 & I\end{array}\right)
\}$
\caption{Expansion of assignment $x = x + bf(cx)$}
\label{fig:concat}
\end{figure}

As an example of how these rules may appear in the code, consider a statement of the form
\verb|x = x + b*f(c*x);|
where the function $f$ is as in lemma \ref{lemma:globlips}. While this statement
does not correspond to
any single rule as written, we can introduce auxiliary \emph{virtual} variables that decompose the
statement into single operations, as shown in figure \ref{fig:concat}.

\section{Synthesizing proofs for autogenerated C code}\label{sec:synthesizing}

The parametric constraints are solved by numerical methods. In particular, we use the Yalmip and
Sedumi packages for Matlab to solve them. Once solved, it
is possible to build a proof of stability of the code with a simple syntax-directed procedure.
This source-to-source translation relies on the fact that the witness provides not symbolic
but actual numerical elements of the abstract domain. Our only task is to replace variables $x$
for which a constraint $x \in D$ is provided with with domain variables $X$ so that the
constraint becomes $X = D$. All other variables are left intact. Thus, we end up with a program
that is a mixture of abstract and concrete variables. We then proceed to execute this program
like abstract interpretation would do. However, since we are executing invariant code, we
are already at the fixpoint, so we do not need to iterate. All that is left to do is
checking that the invariants are valid, and that is a matter of evaluating whether the
corresponding resolved numerical constraints evaluate to true.

An example of the C code and the corresponding Matlab validator is shown in the appendix.

\section{Related work}\label{sec:related}

The study of stability of controllers at the theoretical level is a very mature subject,
especially for linear systems. Two techniques are available in that case: eigenvalue
analysis and Lyapunov methods. While eigenvalue analysis is more straightforward, Lyapunov
methods generalize better to non-linear cases\cite{vidyasagar92nonlinear}. Thus, most recent
work uses the latter. Lyapunov
methods give way to the study of propagation of ellipsoids such as that presented in this paper.

Beyond the theoretical level, several papers have focused on the analysis at the model
level\cite{henzinger97hytech}\cite{alurA3progress}. Among them, the approach most similar to ours
was proposed by Botchkarev et al.\cite{botchkarevA0verification}, who used ellipsoidal calculus
to compute
overapproximations to the reachability set of controller models. They gave algorithms to
estimate unions and intersections of ellipsoids and implemented a tool
called VeriSHIFT to automate the reachability computation of models designed with it.
Their algorithm has been later improved by Casagrande et al.\cite{casagrandeA4improving}.

At the implementation level, Cousot, in a follow-up of some earlier work\cite{cousot78automatic},
showed how to use parametric abstract domains and
external constraint solvers to prove invariance and termination\cite{cousotA5proving}, illustrating
it with several small examples. Simultaneously, Roozbehani et al.\cite{roozbehaniA5modeling}
used similar linear and semi-definite programing to search for Lyapunov invariants for
boundedness and termination.
These methods
have been later extended to prove also non-termination\cite{guptaA8proving}.
Termination has also been studied by Tiwari\cite{tiwariA4termination}, who used eigenvalue
techniques instead of Lyapunov methods to prove termination of linear programs very similar
to those common in linear controllers. Automatic linear invariant generation by means of
non-linear constraint solvers has also been studied by Colon et al.\cite{colonA3linear},
and their work has been generalized to some non-linear invariants by Sankaranarayanan et
al.\cite{sankarA4nonlinear}.

A different route was
followed by Blanchet et al.\cite{blanchetA3static} with the design and implementation of the
heavy-duty ASTREE static analyzer based on abstract interpretation\cite{cousot77abstract}.
The novelty of the ASTREE analyzer is the combination of multiple abstract
domains\cite{cousotA6combination}, such as
rectangular, polyhedral, octagonal and an ad-hoc 2-dimensional version of an ellipsoidal domain,
and its application to large programs.

\section{Conclusion and Future Work}\label{sec:conclusion}

We have shown how to apply the ellipsoidal parametric abstract domain to the analysis of
implementations of controllers. While our focus has been on linear controllers, we have
shown how to extend this analysis to some non-linear cases. Besides proposing some novel treatment
of ellipsoidal relationships, our paper also illustrates how to combine several techniques into
a single methodology for verification of software. We are tempted to call this methodology
\emph{abstract compilation and execution of proofs}
to emphasize both similarity and
distinction from the abstract interpretation methodology. Like abstract interpretation, abstract
compilation performs static analysis based on abstract domains, but instead of evaluating
statements into concrete elements of the abstract domain, we just translate the statement into
constraints in some constraint language. The compiled constraint program is then executed.
Abstract execution takes as input
a candidate witness of constraint satisfaction and produces as output a proof of its validity.
The candidate witness can either be provided by the user or it can be automatically 
discovered by a constraint solver.

We believe that this separation of analysis and synthesis
may provide more modularity for many other verification problems. For example, we did not
incorporate into our analysis a model for floating-point operations, and so the verification
we did is not strictly correct. However, in order to make it correct, we just need to provide
a floating point model to the synthesis phase, because even if the analysis is slightly
incorrect, it would not matter as long as the witness we provide is correct. Providing a
floating point model for the synthesis is a much easier exercise than incorporating the
intricacies of floats into the analysis.

This methodology may also make it easier to use different domains in the analysis and the synthesis
phase. For example, we may explore the use of ellipsoids for the analysis due to its relatively
low computational cost but then convert to the potentially more precise polyhedral domain
in the synthesis phase, thus providing tighter estimates of the behavior. Alternatively, we
may use different domains for the analysis of different parts of a program and then combine them
in the synthesis phase. Ideally, we would like to be able to provide some heuristics for
the automatic selection of suitable abstract domains for different pieces of code. Finally,
we would like to exploit the connection between stability and non-termination, as the
former is just a special case of the latter provided the program fails (and thus terminates)
whenever non-stability (i.e., divergence) is detected. Any correct program intended to run forever
must eventually reach a steady state, and it may be possible to describe this state in terms
of stability with respect to some dynamics. We think this avenue may be worth exploring.

\bibliographystyle{plain}
\bibliography{refs}

\section*{Appendix: example code}

\begin{figure}
\begin{minipage}{0.3\textwidth}
\begin{verbatim}
A = [0.999, 0; 0, 1];
c = [1, 0];
b = [2; 2];
x = [1000; 0];
while 1
  u = fscanf(stdin,"%f");
  x = A*x + b*u;
  y = c*x;
  fprintf(stdout,"%f\n",y);
end
\end{verbatim}
\end{minipage}\hfill
\begin{minipage}{0.6\textwidth}
\begin{verbatim}
#include <stdio.h>

double A[2][2] = { {0.999, 0}, {0, 1} };
double c[2] = { 1, 0 };
double b[2] = { 2, 2 };
double x[2] = { 1000, 0 };
double u,y;

void main(void)
{
  int i,j;
  double x_new[2];

  while(1)
    {
      fscanf(stdin,"%f",&u);
      for(i=0;i<2;i++) {
        x_new[i] = 0;
        for(j=0;j<2;j++) x_new[i] += A[i][j]*x[j];
        x_new[i] += b[i]*u;
      }
      for(i=0;i<2;i++) x[i] = x_new[i];
      y = 0; for(i=0;i<2;i++) y += c[i]*x[i];
      fprintf(stdout,"%f\n",y);
    }
}
\end{verbatim}
\end{minipage}
\caption{A specification (left) and implementation (right) of a linear
system with bounded input}
\label{fig:linbdd}
\end{figure}

\begin{figure}
\begin{verbatim}
x = P; % P is the solution from the constraint solver
while 1,
      u = Q;
      aux = [il1*inv(P),0;0,il2*inv(Q)]; aux=inv(aux); xu = aux;
      for i=1:2, aux = II(x_new)-EE(x_new,i); x_new = aux*x_new*aux';
      end
      for i=1:2,
        for j=1:2,
          aux = [II(x_new), A(i,j)*EE(A,i,j); ZZ(x_new), II(x)];
          aux = aux*[x_new;x]*aux';
          x_new = aux(1:size(x_new),:); x = aux(1+size(x_new),:);
        end
      end
      for i=1:2,
        aux = [II(u), ZZ(x_new); b(i)*EE(b,i), II(x_new)];
        aux = aux*[u;x_new]*aux';
        u = aux(1:size(u),:); x_new = aux(1+size(u),:);
      end
      y = 0;
      for i=1:2,
        aux = [II(y),c(i)*EE(c,i);ZZ(y),II(x)]; aux=aux*[y;x]*aux';
        y = aux(1:size(y),:); x = aux(1+size(y),:);
      end
      if(y < R), % R comes from specs
        break
      end
end
\end{verbatim}
\caption{Propagation rules expressed as a Matlab program}
\label{fig:linbddmat}
\end{figure}

Figures \ref{fig:linbdd} and \ref{fig:linbddmat} show a sample C source and the corresponding
Matlab translation.



\end{document}